\documentclass[twocolumn,showpacs,preprintnumbers,amsmath,amssymb]{revtex4}

\usepackage[dvips]{graphicx}
\usepackage{dcolumn}
\usepackage{bm}

\begin{document}


\title{Nanosecond quantum state detection in a current biased dc SQUID}

\author{J. Claudon$^*$, A. Fay, E. Hoskinson, and O. Buisson}

\affiliation{Institut N\'eel/D\'epartement Nanosciences, CNRS, 25 avenue des Martyrs, BP 166, 38042 Grenoble, France. (*) present adress: CEA-CNRS group "Nanophysique et Semiconducteurs", D\'epartement de Recherche Fondamentale sur la Mati\`ere Condens\'ee, SP2M, CEA Grenoble, 17 avenue des Martyrs, 38054 Grenoble, France, and Institut N\'eel/D\'epartement Nanosciences, CNRS, 25 avenue des Martyrs, BP 166, 38042 Grenoble, France.}

\date{\today}

\begin{abstract}
This article presents our procedure to measure the quantum state of a dc SQUID within a few nanoseconds, using an adiabatic dc flux pulse. Detection of the ground state is governed by standard macroscopic quantum theory (MQT), with a small correction due to residual noise in the bias current. In the two level limit, where the SQUID constitutes a phase qubit, an observed contrast of 0.54 indicates a significant loss in contrast compared to the MQT prediction. It is attributed to spurious depolarization (loss of excited state occupancy) during the leading edge of the adiabatic flux measurement pulse. We give a simple phenomenological relaxation model which is able to predict the observed contrast of multilevel Rabi oscillations for various microwave amplitudes.
\end{abstract}

\pacs{Valid PACS appear here}
\maketitle

\section{Introduction}

The last few years have shown significant breakthrough in the domain of superconducting Josephson qubits. Coherent manipulations have been achieved on a single qubit with various designs \cite{Nakamura_Nature99,Vion_Science02,Martinis_PRL02,Chiorescu_Science03,Wallraff_0405}. Recently,
coupling between two qubits has been demonstrated \cite{Pashkin_Nature03,Berkley_Science03,McDermott_Science05} and quantum logical gates have been realized \cite{Yamamoto_Nature03,Steffen_Science06}. 

The dc SQUID is a versatile quantum circuit displaying several energy levels with tuneable anharmonicity. In the high anharmonicity limit, it becomes a phase qubit whose decoherence has been characterized \cite{Claudon_PRBR06}. With smaller anharmonicity, microwave excitation induces complex dynamics which involve many levels \cite{Claudon_PRL04}.

Readout of the quantum state is one of the key issues in quantum circuits operation. As proposed in \cite{Buisson_PRL03} and experimentally implemented in \cite{Claudon_PRL04}, our measurement of the quantum state of the dc SQUID is performed within a few nanoseconds, using adiabatic dc flux pulses. Depending on the SQUID state, the pulse implements a selective switch to the resistive state. The dc SQUID design thus combines two functionalities: the quantum circuit and the state measurement. A fast measurement allows for a time domain study of fast dynamics. A similar measurement procedure for a phase qubit, developed independently, has been used to investigate the coupling to spurious two level
fluctuators \cite{Cooper_PRL04} and more recently the coupling between two qubits \cite{McDermott_Science05}. Moreover, speed also minimizes contrast losses due to depolarization during the measurement stage. In this article, we present a complete characterization of our measurement performance, both in the two level and multilevel limit.

The paper is organized as follows. Section I introduces the physics of the SQUID and discusses its operation as a quantum circuit. The experimental implementation is presented in Section II, focusing on the microwave part which is a key point of the experimental setup. In the third part, results on the ground state detection are presented. Excepting a small correction due to residual noise on bias parameters, the experiment is consistent with macroscopic quantum tunnelling theory. Detection in the two level limit and measurement in the multilevel limit are then successively presented. A simple relaxation model successfully accounts for measurement performances. To finish, section IV contains concluding remarks and improvement perspectives.

\section{The current bias dc SQUID}

\begin{figure}[t]
\resizebox{0.45\textwidth}{!}{\includegraphics{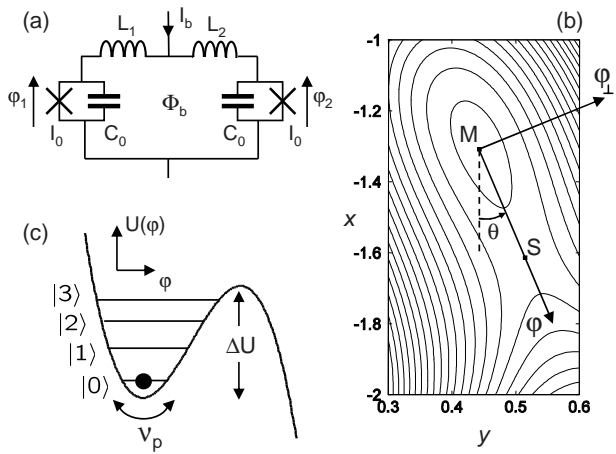}}
\caption{(a) Electrical schematic of the dc SQUID. In the superconducting state, this circuit can be seen as two inductively coupled resonators. The two pure Josephson dipoles, symbolised by crosses, introduce two non-linear inductances controlled by the bias current $I_b$ and the loop current which tends to screen the external applied flux $\Phi_b$. (b) Contour plot of the potential surface $U(x,y)$ versus $x = \tfrac{1}{2}(\varphi_1 + \varphi_2)$ and $y = \tfrac{1}{2}(\varphi_1 - \varphi_2)$, for a typical bias point; $(x)$ and $(y)$ scales are identical, energy scale is arbitrary. Along the escape direction $(\varphi)$, which joins the local minimum M and the only saddle point S, the potential is cubic (c). The quantum states under study are vibration states within the potential well. In the transverse direction $(\varphi^{}_{\perp})$, the potential is roughly quadratic with a bottom frequency which is about twice $\nu_p$. Coupling between the escape and the transverse direction is neglected.}
\label{fig1}
\end{figure}

The dc SQUID, presented in Fig. \ref{fig1}(a), consists of two identical Josephson junctions (JJ), each with a critical current $I_0$ and a capacitance $C_0$. The junctions are embedded in a superconducting loop of total inductance $L_s$ which is unequally distributed between the two arms of the SQUID. The asymmetry in inductance is characterized by the parameter $\eta = (L_1-L_2) /L_s$, where $L_1$ and $L_1$ are respectively the inductance of the first and second arm. In our circuit, the Josephson energy $E_J=(\Phi_0/2\pi) I_0$ is more than 3 orders of magnitude greater than the Cooper pair
Coulomb energy $E_c=(2e)^2/2C_0$ ($\Phi_0 = h/2e$ is the superconducting flux quantum). In this limit, the two superconducting phase differences $\varphi_1$ and $\varphi_2$ across the two Josephson junctions are the natural variables to describe the dynamics of the system. This two-dimensional phase dynamics can be treated as that of a fictitious particle of mass $m = 2C_0(\Phi_0/2\pi)^2$ moving in a two-dimensional potential \cite{Tesche_LTP77}
\begin{equation}
U(x,y)=2 E_J \big[-\big(\tfrac{I_b}{I_0}\big)x-\cos x \cos y - \big(\tfrac{I_b}{I_0}\big) \eta y + b(y- \pi \tfrac{\Phi_b}{\Phi_0})^2 \big],
\end{equation}
where $x=\frac{1}{2}(\varphi_1+\varphi_2)$ and $y=\frac{1}{2}(\varphi_1-\varphi_2)$. The shape of the potential, which contains valleys and mountains, is experimentally controlled through the bias current $I_b$ and the bias flux $\Phi_b$ threading the superconducting loop. Here, the inductance coupling parameter $b=\Phi_0/(2\pi L_s I_0)$ is of order unity. With the values of $I_b$ and $\Phi_b$ used in our experiments, the potential surface displays only one type of local minimum. They are separated from each other by a saddle point, where the particle can escape \cite{Lefevre_PRB92}. A contour plot of the potential surface for a typical bias point is presented in Fig. \ref{fig1}(b). Along the escape direction, which makes an angle $\theta$ with the $x$ axis, the potential is cubic and is completely characterized by its bottom frequency $\nu_p$ and a barrier height $\Delta U$ [Fig. \ref{fig1}(c)]. These two quantities depend on the magnetic flux and vanish at the SQUID's critical current $I_c(\Phi_b)$. We assume complete separation of the variables along the escape direction and the transverse one by neglecting the coupling terms between these two directions. In this approximation, the phase dynamics of the SQUID along the escape direction is similar to the phase dynamics of a current-biased single Josephson junction. The parameters $\Delta U$ and $\nu_p$, explicitly given in the appendix 1, are renormalized, thereby taking into account the two-dimensional nature of the potential.

For bias currents $I_b < I_c$, the particle is trapped in a local minimum and its quantum dynamics is described by the Hamiltonian
\begin{equation}
\hat{H}_0 = \tfrac{1}{2} h \nu_p \big[ \hat{p}_{\varphi}^2+\hat{\varphi}^2 \big]-\sigma h \nu_p \hat{\varphi}^3.
\end{equation}
$\hat{\varphi} = (2 \pi \sqrt{m \nu_p/ h})\hat{\phi}$ represents the
reduced phase operator, associated to the phase $\hat{\phi}=\cos
\theta \hat{x} + \sin \theta \hat{y}$ along the escape
direction. $\hat{p}_{\varphi}$ is the reduced momentum conjugated to
$\hat{\varphi}$. The relative magnitude $\sigma$ of the cubic term
compared to the harmonic term is related to the barrier height $\Delta
U = h \nu_p / 54\sigma^2$. For $I_b$ below $I_c$, several low-lying
quantum states are found near the local minimum. These states,
describing the oscillatory motion within the anharmonic potential, are
denoted $\left| n \right>$ for the $n$th level, with $n= 0, 1, 2, ...$
The corresponding energies $E_n$ were calculated in
Ref. \cite{Larkin_SPJETP86}. The Bohr frequency associated with the
transition $n \rightarrow k$ is denoted $\nu_{n k}$ in the
following. Since energy levels are trapped in a finite height
potential barrier, they are metastable. Escape out of the potential
well is possible both by tunnelling and thermal activation. For
temperatures well below $T^* = h \nu_p/(2 \pi k_B)$, the so-called
cross-over temperature, the thermal process can be neglected
\cite{Balestro_PRL03}. On the timescale of the experiment, only the
three highest energy levels, closest to the top of the barrier, undergo a
significant tunnelling effect. Lower energy levels are stable.

Our procedure to perform quantum experiments consists of the
repetition of an elementary sequence, which is decomposed into four
successive steps. A bias current $I_b$ is first switched on through
the SQUID at fixed magnetic flux $\Phi_b$. The working point
$(I_b,\Phi_b)$ defines the geometry of the potential well. The $I_b$
risetime is slow enough to induce adiabatic transformation of the
potential and the circuit is initially in the ground state
$\left|0\right>$. A microwave flux pulse characterized by its
frequency $\nu$ and amplitude $\Phi_{\mu w}$ manipulates the quantum
state of the system. It induces a time-dependent perturbation term $-
h a_{\mu w} \sqrt{2} \cos(2 \pi \nu t) \hat{\varphi}$ in the Hamiltonian
which couples neighbouring levels. In this
expression, $a_{\mu w}$ is directly proportional to $\Phi_{\mu w}$ and
corresponds to the frequency of the Rabi oscillation between
$\left|0\right>$ and $\left|1\right>$ for an excitation tuned on
$\nu_{01}$ (see appendix 2). Then, a dc flux pulse of amplitude
$\Phi_m$ is applied and brings the system to the measuring point
$(I_b,\Phi_b+\Phi_m)$. This flux pulse adiabatically reduces the
barrier height and allows tunnelling escape of the localised states to
finite voltage states during a time $\Delta t$ of the order of a few
nanoseconds. As discussed in Ref. \cite{Buisson_PRL03}, with precise
adjustment of $\Phi_m$ it is theoretically possible to induce a
selective escape of excited states. Because the SQUID is hysteretic,
the zero and finite voltage states are stable and the result of the
measurement can be readout by monitoring the voltage $V_s$ across the
dc SQUID. Escape out of the well corresponds to the detection of a
voltage which is twice the superconducting gap of the circuit
material. The current bias $I_b$ must be switched off to reset the circuit in the ground state. This
elementary sequence is repeated to extract the occupancies of the excited states.

This article is devoted to the characterisation of the nanosecond
measurement procedure. The experiments presented here only involve
levels $\left|0\right>$ to $\left|3\right>$. Denoting $p_n$ the
population of the $n$th level, we thus restrict discussion to states
characterized by the set of occupancies $\{p_0, p_1, p_2, p_3\}$;
generalization to higher energy levels is straightforward. For such a
state, the escape probability $P_e$ out of the potential well reads:
\begin{equation}
P_e = P_e^0 + \sum_{n=1}^3 (P_e^n-P_e^0) \times p_n.
\end{equation}
$P_e^n(\Phi_m)=1-\exp [-\Gamma_n(\Phi_m) \Delta t ]$ is the escape probability of the state $\left|n\right>$. From the measurement point of view, $P_e^n$ correponds to the probability to detect this level when it is fully populated. The tunnelling escape rate $\Gamma_n$ depends on the geometry of the potential barrier which is in turn fixed by the measuring point $(I_b,\Phi_b+\Phi_m)$. Escape rates are obtained from semiclassical calculations. Particularly, $\Gamma_0$ is given by the usual expression for macroscopic quantum tunnelling of the ground state of a single Josephson junction,
\begin{equation}
\Gamma_0 = 12 \sqrt{6 \pi \tfrac{\Delta U}{h \nu_p}} \nu_p \exp (-\tfrac{36}{5} \tfrac{\Delta U}{h \nu_p}),
\label{eq:MQT}
\end{equation}
and the contrast in escape rate between two successive levels is
$\Gamma_{n+1}/\Gamma_n = \exp(2\pi) \approx 535$
\cite{Larkin_SPJETP86}. In the following, we refer to the plot of
$P_e^n$ versus $\Phi_m$ as the escape curve of level $\left| n \right>$.

\begin{figure}
\resizebox{0.45\textwidth}{!}{\includegraphics{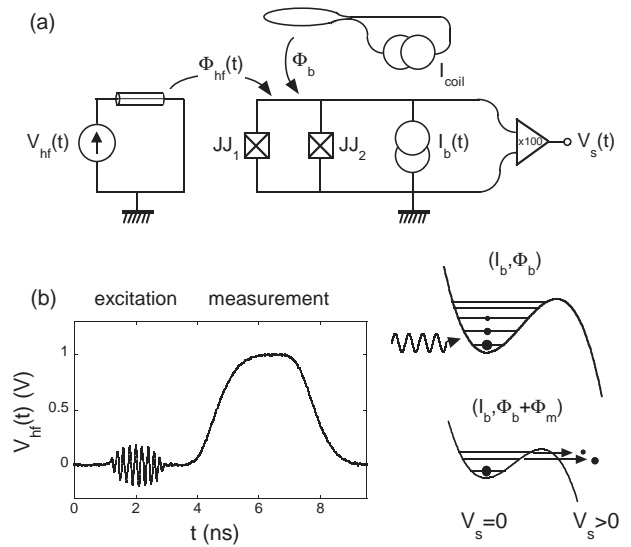}}
\caption{Principle of the SQUID operation. (a) Schematic of the
  electrical circuit. The capacitively shunted JJs are symbolised by a
  square with diagonals. In our experiments, $I_b$ is first switched
  on through the SQUID at fixed magnetic flux $\Phi_b$, defining the
  shape of the potential well. A high frequency flux signal is then
  applied on the superconducting loop with a microwave antenna. A
  digital sampling oscilloscope record of this signal is presented in
  (b). It is composed of a microwave pulse which induces transition
  between adjacent energy levels, followed by a measuring dc pulse
  which adiabatically reduces the barrier height and induces a
  selective tunnelling escape of excited states. Escape is detected by
  a voltage $V_s$ across the circuit, amplified at
  room-temperature. At the end of the sequence, $I_b$ is switched off
  and the circuit is reset to the superconducting ground state.}
\label{fig2}
\end{figure}

In the two level limit, the detection contrast between states $\left|
  0 \right>$ and $\left| 1 \right>$ is $P_e^1-P_e^0$. The measuring
amplitude which optimizes the detection probability allows a detection
contrast of $0.98$ and corresponds to $P_e^0 = 0.02$. With these
settings, in the general case where more than two levels are populated
in the system, the escape probability is practically equal to the
population of excited states:
\begin{equation}
P_e = 0.02 + 0.98 p_1 + p_2 + p_3.
\end{equation}
Moreover, it is possible to find a lower pulse amplitude for which the
ground and first excited states remain trapped, but states $\left| 2
\right>$ and $\left| 3 \right>$ escape with close to 100\%
probability. Repeating this procedure with successively smaller pulse
amplitudes, one can measure $p_1 + p_2 + p_3$, $p_2 + p_3$ and $p_3$
and therefore reconstruct all levels occupancies.

Our measurement is one of the fastest implemented in Josephson quantum circuits. The rise time of the measuring pulse is $1.6 \: \text{ns}$ and the duration $\Delta t$ can be as short as $1.5 \: \text{ns}$. This speed allows a time domain study of fast relaxing phenomena and tends to limit the loss of measurement contrast due to relaxation. The question of the possibility of nonadiabatic effects, namely transitions between energy levels which modify the levels occupancies, has been recently addressed for a rf SQUID which is a very similar device \cite{Zhang_CondMat06}. Calculations show that the adiabatic error is negligible for the pulses used in our experiments: it is less than $10^{-3}$ for $\Delta t = 1.5 \: \text{ns}$ and decreases very rapidly for slower pulses.

In terms of speed, escape to the running state of a tilted washboard potential may also have an advantage over the method used for the rf SQUID \cite{Cooper_PRL04}. In that case, the fictitious particle escapes to another trapped state, which can interact strongly with the qubit levels if relaxation is insufficient. This leads to retraping of the particle in the qubit states. This has been investigated numerically in Ref. \cite{Zhang_CondMat06}. Assuming an infinite $T_1$, the retrapping probability was found to be greater that $15\: \%$. In our case, the qubit excited state escapes to an unbound state of the tilted washboard. Retrapping is therefore much less likely, and faster measurement times possible.


\section{Experimental setup}

\begin{figure}
  \resizebox{0.35\textwidth}{!}{\includegraphics{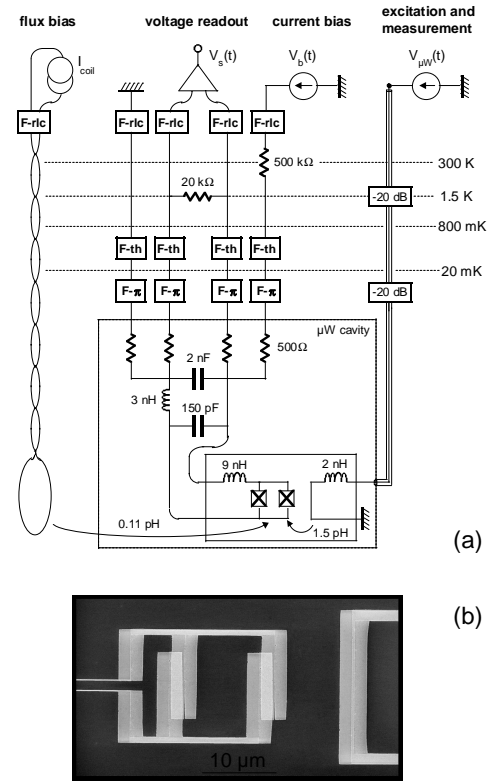}}
  \caption{(a) Schematics of the experimental setup with temperatures
    indicated on the right. \fbox{{\footnotesize F-rlc}} stands for a
    second order low-pass rlc filter, \fbox{{\footnotesize F-th}} is a
    1 m lossy thermocoax and \fbox{{\footnotesize F-$\pi$}} is
    composed of two cascaded $\pi$ filters and two meters of lossy
    thermocoax. \fbox{{\footnotesize -20 dB}} is a $50\: \Omega$
    attenuator. (b) Scanning electron microscope picture of the
    SQUID. The circuit is obtained by shadow evaporation of two $15\:
    \text{nm}$ aluminium layers; Josephson junctions appear in light
    grey. The extremity of the $500\: \mu\text{m}$ long thin current
    bias line is visible on the left: their width is between $200\:
    \text{nm}$ and $300\: \text{nm}$. The end of the on-chip $\mu$w
    antenna appears on the right.}
\label{fig3}
\end{figure}

The measured circuit was realized using e-beam lithography and shadow
evaporation of aluminium on an oxidized silicon chip. A SEM picture
of the circuit is presented in Fig. \ref{fig3}(b). The SQUID consists
of two large JJs of $15\: \mu\text{m}^2$ area enclosing a $350\:
\mu\text{m}^2$-area superconducting loop. Each Al/AlOx/Al junction has
a critical current $I_0=1.242\: \mu\text{A}$ and a capacitance
$C_0=0.56\: \text{pF}$. The two SQUID branches combine to give a total
loop inductance $L_s=280\: \text{pH}$ with asymmetry parameter $\eta =
0.414$. These electrical parameters are obtained by combining the
analysis of two independent measurements. First, $I_0$, $L_s$, $\eta$
and an approximate value of $C_0$ are extracted from macroscopic
quantum tunnelling measurements, as discussed in
Ref. \cite{Balestro_PRL03,Claudon_PRL04}. Second, spectroscopic
measurements of the $\left| 0
\right> \rightarrow \left| 1 \right>$ transition frequency $\nu_{01}$ as a function of $I_b$ are used to determine a precise
value for $C_0$ \cite{Claudon_PRL04,Claudon_PRBR06}.  All measurements are performed with the sample at a
temperature of $30\: \text{mK}$ in a dilution fridge. Thermal
energy is then small compared to the relevant energy scales of the
circuit, namely the Josephson energy $E_J=k_B \times 29.6\: \text{K}$
and the vibration quantum $\sqrt{E_J E_c}=k_B \times 442\: \text{mK}$,
with $k_B$ the Boltzmann constant.

To avoid spurious microwave resonances, the chip is mounted in a
shielded copper cavity whose cut-off frequency is above 20 GHz. The
immediate electromagnetic environment of the SQUID, located in the
microwave cavity, is designed to decouple the circuit from the
external room temperature classical electrical apparatus. It consists
of two cascaded filters. The first and most important includes a $9\:
\text{nH}$ on-chip inductance introduced by long on-chip
superconducting thin wires, and a gold thin film $150\: \text{pF}$
parallel capacitor. The second filter consists of the bonding wires, with an estimated
inductance $3\: \text{nH}$, and a surface mounted $2\: \text{nF}$
capacitor. This electrical environment displays two low frequency
resonances whose measured frequencies are consistent with the above parameters \cite{These_Claudon}. As
illustrated on Fig. \ref{fig3}(a), special care was taken in bias line
filtering, and the experimental setup includes several low temperature
filters. Bias flux is applied with a copper coil cooled down to $30\:
\text{mK}$. External low frequency flux fluctuations are screened with
soft iron and mu-metal at room temperature and a superconducting lead
shield at $1.5\: \text{K}$. The room temperature high frequency signal
is guided by $50\: \Omega$ coax lines and attenuated twice by 20 dB at
$1.5\: \text{K}$ and $30\: \text{mK}$, before reaching the SQUID
through an on-chip short circuit. The antenna is located $10
\mu\text{m}$ from the SQUID, leading to a $1.5\: \text{pH}$ mutual
inductance coupling. The $\mu$w line is terminated by an inductance
estimated to be $2\: \text{nH}$ which introduces a supplementary $6\:
\text{dB}$ attenuation at $10\: \text{Ghz}$. Special care was taken to
minimize the coupling between the high frequency antenna and the
bonding wire loops in the direct electrical environment of the SQUID.

The high frequency signal used in our experiments results from the
combination of a microwave excitation pulse and a measurement dc
pulse. The $\mu$w pulse is generated by mixing continuous microwaves
with a shaping dc pulse. We define the risetime as the duration to
reach $95\: \%$ of the peak amplitude starting from $5\: \%$. The
duration is defined as the time interval where the pulse amplitude is
at least $95\: \%$ of the maximal amplitude. With this convention, the
shaping pulse has a $0.8\: \text{ns}$ risetime and a duration tuneable
from $1\: \text{ns}$ to several milliseconds. The measurement pulse
has a $1.6\: \text{ns}$ risetime and a minimum duration of $1.6\:
\text{ns}$. Both microwave shaping and measurement pulse are produced
by two outputs of one dc pulse generator, allowing precise control of
timing. In Fig. \ref{fig2}(c), a digital sampling oscilloscope record
of the signal is presented. As a demonstration of signal generation
performance, the duration of the microwave and measurement pulse have
been reduced to the minimum: the whole sequence is then finished in
less than $7\: \text{ns}$. Excepting specific measurements, the delay
between microwave and measurement is kept as short as $1\: \text{ns}$
to minimize depolarization due to population relaxation. The microwave
amplitude $a_{\mu w}$ can be tuned from $0$ to $300 \:
\text{MHz}$. Depending on the bias point, the measuring pulse
amplitude ranges between a few $10^{-2} \: \Phi_0$ and $2 \times
10^{-1} \: \Phi_0$.

To limit heating in the SQUID voltage state, a specific electronic circuit cuts the bias current to zero as soon as a finite voltage across the SQUID is detected. The voltage readout time, about $10 \: \mu \text{s}$, is determined by the bandwidth of the strongly filtered voltage lines. With a wait of 1 ms after each voltage measurement to allow the sample to cool back down to base temperature, the overall repetition rate is about $1\: \text{kHz}$. The experiment sequence is
repeated up to 5000 times such that the statistical noise on $P_e$ has a theoretical standard deviation lower than $4 \times 10^{-3}$.

\section{Experimental results}

In this section, we present an experimental characterization of the detection performances. All the results have been obtained at the working point $W_p =(2.222\: \mu\text{A}, -0.117\: \Phi_0)$. The bias flux is close to the one which maximizes the critical current (see inset in Fig. \ref{fig4}) and $W_p$ thus presents a low sensitivity to flux fluctuations. Using the electrical parameters of the SQUID, the characteristics of the potential well are calculated: $\nu_p=8.428\: \text{GHz}$ and $\Delta U = h \times 71.1\: \text{GHz}$. The well contains 8 energy levels and the anharmonicity is $\nu_{01}-\nu_{12}=160\: \text{MHz}$. The measuring pulse has a duration $\Delta t = 4 \: \text{ns}$. Our setup allows for faster measurement, but this timing preserves a trapezoidal pulse shape with a well defined plateau. This simple shape makes the application of tunnelling theory straightforward.

\subsection{Detection of the ground state}

\begin{figure}
\resizebox{0.45\textwidth}{!}{\includegraphics{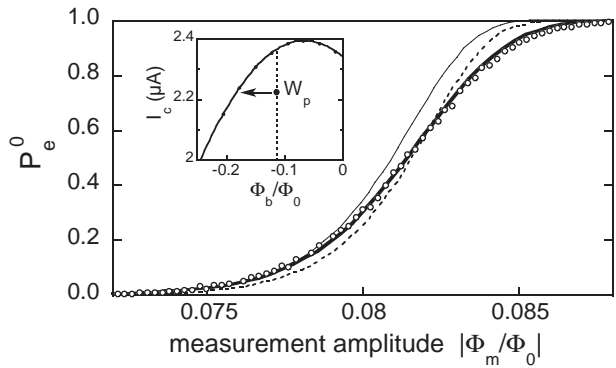}}
\caption{Detection probability of the ground state versus the amplitude of a $4 \: \text{ns}$ measuring pulse. The fit of experimental data ($\circ$) to pure tunnelling theory (dashed line) is in qualitative agreement with experiment. A closer fit is achieved by taking into account a low frequency current noise with r.m.s value $\big< \delta I^2 \big>{}^{1/2} = 5\: \text{nA}$, which broadens the escape curve (thick solid line). The effect of a high frequency noise with the same r.m.s fluctuations is also plotted (thin solid line). Inset: the working point $W_p$ in the bias parameters space. A negative flux pulse brings the system close to the critical line $I_c(\Phi_b)$.}
\label{fig4}
\end{figure}

Fig. \ref{fig4} shows a plot of the experimental detection probability $P_e^0$ of the ground state versus the measuring pulse amplitude. The measured curve is slightly wider than the MQT prediction (Eq. \ref{eq:MQT}). To quantify this, we introduce the width $\Delta \Phi$ of the transition, defined as the flux range between $P_e^0=0.05$ and $P_e^0=0.95$. MQT predicts $\Delta \Phi = 7.46 \times 10^{-3}\: \Phi_0$ whereas the experimental value corresponds to $9.52 \times 10^{-3}\: \Phi_0$. As discussed below, this broadening is an effect of residual noise in the bias parameters.

We restrict the discussion to adiabatic noise, whose frequencies are below the typical transition frequency of the SQUID, approximately given by $\nu_p$. Such a perturbation does not induce transitions between levels. Without any loss of generality, we can focus on a current noise $\delta I$. The comparison between the fluctuation frequency and the timescale of the experiment plays a crucial role on the impact of noise. Two limiting situations can be distinguished. Fluctuations much slower than $\Delta t^{-1}$ are named low frequency noise. In this case, $\Gamma_0$ is constant during one measurement pulse, but randomly fluctuates from one measurement to the other. Assuming that 5000 repetitions is enough to explore all realizations of $\delta I$ (ergodicity hypothesis), the actual measured probability is the average
\begin{equation}
\left< P_e^{n} \right>_{lf} = 1 - \big< \exp \big[ -\Gamma_n(I_b+\delta I,\Phi_b+\Phi_m) \Delta t \big] \big>,
\end{equation}
where the brackets denote an averaging over the different values of $\delta I$. As illustrated in Fig. \ref{fig4}, this averaging broadens escapes curves. On the other hand, high frequency noise, with fluctuations much faster than $\Delta t^{-1}$, induce random fluctuations of escape rate during one measuring pulse. Following Ref. \cite{Martinis_PRB88}, the corresponding high frequency averaging reads
\begin{equation}
\left< P_e^{n} \right>_{hf} = 1 - \exp \big[ -\big< \Gamma_n(I_b+\delta I,\Phi_b+\Phi_m) \big> \Delta t \big].
\end{equation}
At lowest order, it just translates the escape curve (see Fig. \ref{fig4}) and does not modify its shape.

Thanks to the massive filtering and shielding of the experimental setup, noise on bias parameters is mostly generated by the close environment of the SQUID which is thermalized at $30\: \text{mK}$. The associated thermal frequency $\nu_T=k_B T/h = 600\: \text{MHz}$ is much lower than the typical transition frequency $\nu_p \approx 8 \: \text{GHz}$. Therefore, the noise can be treated as an adiabatic perturbation. 

In Ref. \cite{Claudon_PRBR06}, escape measurements of the ground state induced by slow current pulses exhibited a low frequency flux noise for the same experimental setup and SQUID. Its r.m.s. value $\big< \delta \Phi^2\big>{}^{1/2} = 5.5 \times 10^{-4}\: \Phi_0$, corresponding to noise frequencies ranging from $100 \: \text{mHz}$ to $20\: \text{kHz}$, is largely  insufficient to account for the broadening of the width. In the following, this flux noise is neglected. 

The close electrical environment of the SQUID also generates Nyquist current fluctuations whose spectrum density is easily derived from the quantum fluctuation-dissipation theorem. Integration between $100 \: \text{mHz}$, the inverse of a 5000 repetition sequence duration, and $\Delta{t}^{-1} = 250 \: \text{MHz}$ gives $\big< \delta I^2\big>{}^{1/2} = 5.5 \: \text{nA}$. The best fit to experimental data, presented in Fig. \ref{fig4}, is obtained when a low frequency Gaussian current noise with $\big< \delta I^2\big>{}^{1/2} = 5 \: \text{nA}$ is taken into account. This value is in good agreement with the previous calculation and the whole analysis demonstrates we control the measurement of the ground state on the nanosecond scale.

\subsection{Detection in the two level limit}


In the two level limit, the escape probability takes the simple form:
\begin{equation}
P_e = \left<P_e^0\right> + \big[ \left<P_e^1\right> - \left<P_e^0\right> \big] \times p_1.
\label{eq:echap2}
\end{equation}
The two detection probabilities $\left<P_e^0\right>$ and $\left<P_e^1\right>$ depend on $\Phi_m$. They are calculated taking into account the low frequency current noise previously discussed (for clarity, the index ''$_{lf}$'' is implicit). This noise slightly broadens escape curves and thus affects the contrast of detection. The expected contrast $\left< P_e^1 \right>-\left< P_e^0 \right>$ is 0.95 (compared to the 0.98 predicted by the pure MQT regime). We note that high frequency noise does not change the separation between escape curves of states $\left| 0 \right>$ and $\left| 1 \right>$ and leaves the contrast of detection unchanged.

\begin{figure}
\resizebox{0.45\textwidth}{!}{\includegraphics{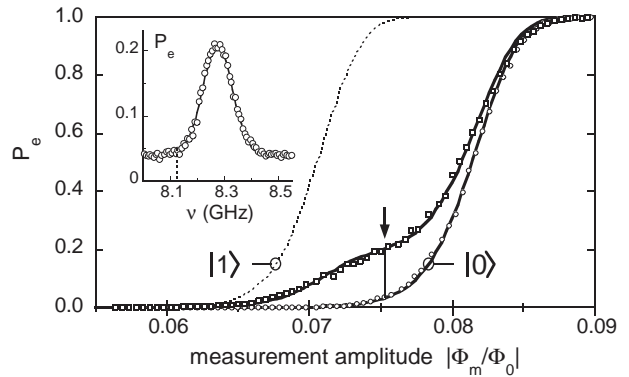}}
\caption{Detection of level $\left| 1 \right>$ partially populated with a $a_{\mu w}=16.4\: \text{MHz}$ amplitude microwave pulse ({\tiny $\square$}). The solid line is a fit with $p_1$ as the only free parameter, giving $p_1=0.18$. The experimental detection probability of the ground state ($\circ$), and the calculated detection probability of $\left| 1 \right>$ (dashed line) are also plotted. The optimal detection contrast, pointed to by an arrow, is obtained for $\Phi_m = -0.0754\: \Phi_0$; this setting corresponds to $P_e^0 = 0.03$. Inset: spectroscopy of the $\left| 0 \right> \rightarrow \left| 1 \right>$ transition for the same $\mu$w amplitude. The circles are experimental data and the continuous line is a fit to a Gaussian lineshape. The resonance peak is centred on $\nu_{01}=8.283\: \text{GHz}$ with a $110\: \text{MHz}$ full width at half maximum.}
\label{fig5}
\end{figure}

In the following, $\left| 1 \right>$ is first populated with a low power microwave pulse whose $300\: \text{ns}$ duration is sufficient to reach a stationary state where $p_1$ depends only on $\nu$ and $a_{\mu w}$. The inset of Fig. \ref{fig5} shows the resonance curve associated with the $\left| 0 \right> \rightarrow \left| 1 \right>$ transition for a microwave amplitude $a_{\mu w}=16.4\: \text{MHz}$. A fit to a Gaussian lineshape leads to $\nu_{01}=8.283\: \text{GHz}$ and a full width at half maximum equal to $110\: \text{MHz}$. The location of the $\left| 1 \right> \rightarrow \left| 2 \right>$ transition, with resonant frequency $\nu_{12} = \nu_{01} - 160 \: \text{MHz}$, is indicated by the vertical dashed line to the left of $\nu_{01}$ in the inset. This transition is not excited for such a microwave amplitude. 

The main plot of Fig. \ref{fig5} shows the escape probability after a microwave pulse tuned to $\nu_{01}$, versus the measurement amplitude $\Phi_m$. The experimental detection probability of the ground state, already discussed, is also plotted. The contrast of the detection is directly proportional to the difference of these two curves. To help interpretation, the calculated detection probability of $\left| 1 \right>$ is also shown. For small $\Phi_m$, $\left| 0 \right>$ and $\left| 1 \right>$ remain trapped within the well. When $\Phi_m$ is increased, a clear bulge appears: it
corresponds to the selective escape of state $\left| 1 \right>$. At higher $\Phi_m$, the potential barrier is so small that both levels escape. A fit to Eq. \ref{eq:echap2} quantitatively reproduces the data and gives $p_1=0.18$. The experimental optimization of the contrast leads to the amplitude $\Phi_m = -0.0754\: \Phi_0$ which corresponds to a ground state detection probability of about $0.03$, a value consistent with theory. In the next paragraph, the measurement amplitude is tuned on this optimal value.

\

\begin{figure}
\resizebox{0.45\textwidth}{!}{\includegraphics{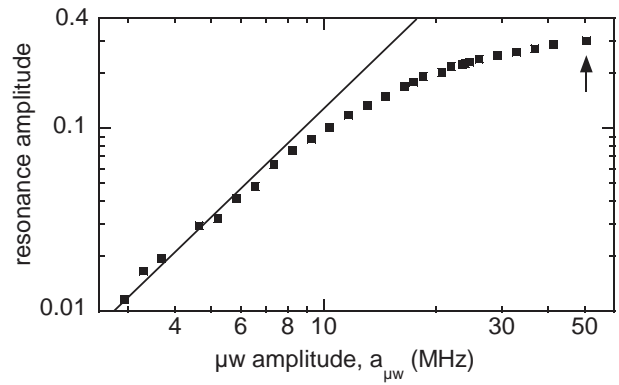}}
\caption{Amplitude of the resonance peak, such as shown in the inset of Fig. \ref{fig5}, versus the $\mu$w amplitude plotted with a double logarithmic scale. The measurement amplitude has been tuned to maximize the detection contrast. Well below a saturation $\mu$w amplitude about $15 \: \text{MHz}$, the experimental data ({\tiny $\blacksquare$}) display a square law dependence (solid line). In the  high $\mu$w amplitude regime the $\left| 0 \right> \rightarrow \left| 1 \right>$ transition is saturated.}
\label{figsat}
\end{figure}


To quantify the detection contrast, we increased the $\mu$w amplitude to fully saturate the $\left| 0 \right> \rightarrow \left| 1 \right>$ transition. The ampitude of the resonance peak versus the $\mu$w amplitude is plotted in Fig. \ref{figsat}. The experimental data display two regimes separated by a saturation $\mu$w amplitude about $15 \: \text{MHz}$. The low $\mu$w amplitude regime is characterized by a square law dependence (solid line). Well above $15 \: \text{MHz}$, the resonance amplitude is nearly independent on the $\mu$w amplitude : the $\left| 0 \right> \rightarrow \left| 1 \right>$ transition is saturated. We now focuss on the measurement corresponding to $a_{\mu w}=50.5\: \text{MHz}$, with an associated resonance amplitude equal to $0.30$. Unfortunately, for such a high power, the $\mu$w excitation is less selective and a slight population of $\left| 2 \right>$ occurs. A rough estimation using multilevel Rabi theory \cite{Claudon_PRL04} gives for this contamination $p_2=0.06$. The corrected experimental population of $\left| 1 \right>$ is thus $0.30 - 0.06 = 0.24$. The theoretical value is $0.50-0.06=0.44$. The ratio of these two quantities, equal to $0.54$, is the optimal contrast of the detection. It is significantly smaller than the expected value of $0.95$. To understand the reason for this contrast loss, we investigate in the next paragraph the effects of relaxation during the measurement flux pulse.\\


Relaxation can occur at three different stages of the experiment: between the $\mu$w pulse and the flux ramping of the measurement pulse (1), during the flux ramping (2) and during the measurement stage (3).
\begin{enumerate}
\item At the working point $W_p$, the energy relaxation of state $\left| 1 \right>$ has been experimentally investigated using $T_1$ NMR-like measurements. $\left| 1 \right>$ is first selectively populated with a $300\: \text{ns}$ microwave pulse tuned at $\nu_{01}$, with an amplitude $a_{\mu w} = 16.4\: \text{MHz}$. Then the state occupancy is measured with increasing time delay after the end of the $\mu$w pulse. The escape probability follows an exponential relaxation with a rate $1/\Gamma_R^1 = 60\: \text{ns}$ (not presented here, see Ref. \cite{Claudon_PRBR06} for a similar curve). Here the delay between flux ramping and $\mu$w is kept as short as $1\: \text{ns}$ and the resulting depolarization is $1.6\: \%$.

\item During the adiabatic flux ramp, $\nu_{01}$ decreases from $8.28\: \text{GHz}$ to $5.42\: \text{GHz}$. During this $2.8\: \text{GHz}$ frequency sweep, the SQUID can couple to spurious resonators located in the dielectric tunnel barrier of the junctions, as demonstrated in Ref. \cite{Cooper_PRL04,Martinis_PRL05}.

\item Finally, at the top of the flux pulse, relaxation competes with tunnelling, yielding a contrast loss given by the ratio $\Gamma_R^1/\Gamma_1$ of these two processe rates. A fast measurement implies high tunnelling rate and thus low
  depolarization. Indeed, at the measuring point, $\Gamma_1$ is typically $2.5 \: \text{GHz}$. An extrapolation of the results presented in Ref. \cite{Claudon_PRBR06} to the transition frequency $\nu_{01}=5.42\: \text{GHz}$ suggests a relaxation rate of about $33\: \text{MHz}$. Therefore, the contrast loss is very small, about $1.3\: \%$.
\end{enumerate}
In the following, we neglect depolarization before and after the flux ramping. We assume that depolarization during the flux ramping is entirely responsible for the observed contrast loss. We describe this process with the help of an effective relaxation rate $\Gamma_{\mathrm{eff}}$ from $\left| 0 \right>$ to $\left| 1 \right>$ which acts during the flux ramping time $\tau=1.6 \: \text{ns}$. The population $p_1(\tau)$ just after the ramp is related to the population $p_1^0$ after the $\mu$w pulse by $p_1(\tau)=p_1^0 \exp(-\Gamma_{\mathrm{eff}} \tau)$. $\Gamma_{\mathrm{eff}}$, the only free parameter of this phenomenological model, is deduced from the experimental detection contrast: $0.95 \times \exp(-\Gamma_{\mathrm{eff}}
\tau)=0.54$ gives $\Gamma_{\mathrm{eff}}=350\: \text{MHz}$. In the next section, we show that this model can be extended to account for measurement performance when more than 2 levels are populated. We will distinguish between $\{p_n^0\}$, the set of occupancies just after $\mu$w and $\{p_n(\tau)\}$, the level populations just after the end of the flux ramping.

\subsection{Multilevel detection}

\begin{figure}
\resizebox{0.45\textwidth}{!}{\includegraphics{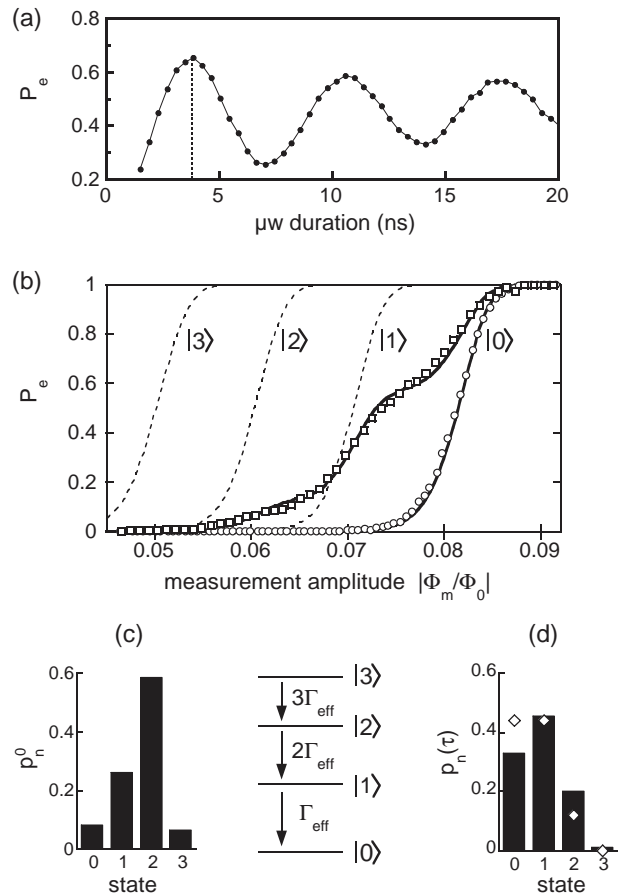}}
\caption{(a) Experimental Rabi oscillation for $a_{\mu w}=164.4\: \text{MHz}$. (b) Detection of excited states populated by a Rabi pulse with duration $t_{\mu w}=3.8\: \text{ns}$. The solid line is a theoretical adjustment of experimental data ({\tiny $\square$}). For comparison, the experimental detection probability of $\left| 0 \right>$ is plotted ($\circ$). The calculated detection efficiencies of pure states $\left| 1 \right>$ to $\left| 3 \right>$ are also plotted as dashed lines. (c) Calculated population distribution after the end of $\mu$w excitation, neglecting decoherence. During the flux ramping, this distribution relaxes to the calculated configuration plotted as solid bars in (d). The diamonds are the experimental populations extracted from the fit of curve (b).}
\label{fig6}
\end{figure}

\begin{figure}
\resizebox{0.4\textwidth}{!}{\includegraphics{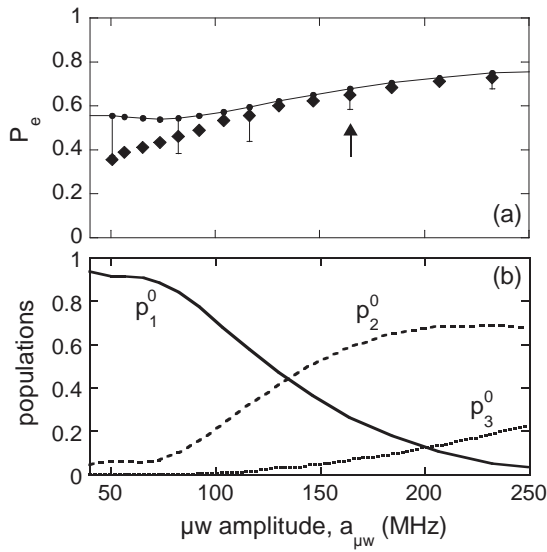}}
\caption{(a) Measured escape probability after a Rabi $\pi$ pulse versus microwave amplitude. The measurement pulse amplitude has been set to maximize the detection contrast between $\left| 0 \right>$ and $\left| 1 \right>$. The solid line with solid circles is a calculation with no free parameter (details in the text) consistent with experimental data ({\footnotesize $\blacklozenge$}). (b) Set of calculated populations after the microwave pulse (without decoherence).}
\label{fig7}
\end{figure}

To investigate the measurement performances in the multilevel regime, the states $\left| 1 \right>$, $\left| 2 \right>$ and $\left| 3 \right>$ are populated with a Rabi pulse. Fig. \ref{fig6}(a) presents a Rabi oscillation induced by microwaves tuned on $\nu_{01}$ with an amplitude $a_{\mu w}=164.4\: \text{MHz}$ similar to the anharmonicity of the potential well. The oscillation is damped in time with the characteristic time $\tilde{T}_2=19\: \text{ns}$. 

A microwave duration $t_{\mu w}=3.8\: \text{ns}$ maximizes $P_e$. In the two level limit, this setting corresponds to a Rabi $\pi$ pulse, which coherently brings the system from $\left| 0 \right>$ to $\left| 1 \right>$. With the help of the model developed in Ref. \cite{Claudon_PRL04}, the population distribution $\{ p_n^0 \}$ in the present case is calculated (see Fig. \ref{fig6}(c)). Particularly, the population of $\left| 2 \right>$ nearly reaches $0.60$ and the higher energy state $\left| 3 \right>$ is occupied at about $0.10$. Fig. \ref{fig6}(b) shows the evolution of the escape probability versus $\Phi_m$. To clarify the interpretation, the calculated escape curves of levels $\left| 0 \right>$ to $\left| 3 \right>$ are also plotted. Surprisingly, there is practically no detectable escape of $\left| 3 \right>$. The first noticeable bulge corresponds to the selective escape of $\left| 2 \right>$ and the second one to the escape of $\left| 2 \right>$ and $\left| 1 \right>$. For higher measurement amplitudes, all the states escape. The experimental occupancies $\{p_n(\tau)\}$ just after the flux ramping can be extracted from a fit to the following formula:
\begin{equation}
P_e = \left<P_e^0\right> + \sum_{n=1}^3 \big[ \left<P_e^n\right> - \left<P_e^0\right> \big] \times p_n(\tau),
\end{equation}
where the escape probabilities $\left< P_e^n \right>$ depends on $\Phi_m$. From the fit, plotted as a solid curve in Fig. \ref{fig6}(b), we obtain $p_1(\tau)=0.44$, $p_2(\tau)=0.12$ and $p_3(\tau) \approx 0$. This set of occupancies, plotted as diamonds in Fig. \ref{fig6}(d), is completely different from the predicted values after the $\mu$w Rabi pulse. Such a striking discrepancy can be explained by a relaxation cascade during the flux ramping. In the following, we expand the relaxation model proposed in the previous section to a 4 level system. 

We keep for $\left| 1 \right>$ the relaxation rate $\Gamma_{\mathrm{eff}}$. As $n$ increases, the relaxation rate of level $\left| n \right>$ dramatically increases. The relaxation rate associated with the random transition $\left| n \right>$ to $\left| n-1 \right>$ is proportional to $\big| \big< n \big| \hat{\varphi} \big| n-1 \big> \big|{}^{-2} = n$. If we neglect other relaxation channels, such as $\left| n \right> \rightarrow \left| n-2 \right>$, with a relaxation rate proportional to the anharmonicity coefficient $\sigma$, the relaxation rate of the $n$th level is $n \Gamma_{\mathrm{eff}}$. Thus, populations evolve in time according to the closed set of master equations:
%
%
\begin{equation}
\frac{d}{dt} \left( \begin{array}{c}
p_3\\
p_2\\
p_1\\
p_0\\
\end{array} \right) =
\left( \begin{array}{c c c c}
-3\Gamma_{\mathrm{eff}} & 0                       & 0                      & 0 \\
+3\Gamma_{\mathrm{eff}} & -2\Gamma_{\mathrm{eff}} & 0                      & 0 \\
0                       & +2\Gamma_{\mathrm{eff}} & -\Gamma_{\mathrm{eff}} & 0 \\
0                       & 0                       & +\Gamma_{\mathrm{eff}} & 0 \\
\end{array} \right) \times
\left( \begin{array}{c}
p_3\\
p_2\\
p_1\\
p_0\\
\end{array} \right).
\end{equation}
which is easily solved analytically. The populations $\{p_n(\tau)\}$ are obtained starting from the initial populations $\{p_n^0\}$ after $\mu$w excitation and letting relaxation act during the ramping time $\tau$. The result, plotted as solid bars in Fig. \ref{fig6}(d) is in relatively good agreement with experimental values. However, our model still overestimates $p_2(\tau)$. This may be due to the neglected $\left| 2 \right> \rightarrow \left| 0 \right>$ relaxation channel or to decoherence. Nevertheless, the essential of the physics is captured by this simple model.

Fig. \ref{fig7}(a) presents the measured escape probability after a Rabi $\pi$ pulse with increasing microwave
amplitudes. In this experiment, the measurement pulse amplitude has been set to maximize the detection contrast between $\left| 0 \right>$ and $\left| 1 \right>$. Our phenomenological model can be used to calculate the global detection efficiencies of each level and to predict the final escape probability given an arbitrary initial population distribution $\{p_n^0\}$ {\it before} the measurement pulse:
\begin{equation}
P_e = 0.03 + 0.54 p_1^0 + 0.79 p_2^0 + 0.90 p_3^0.
\end{equation}
Without any free parameter, this expression describes very well the measured escape probability obtained with high $\mu $w amplitude. Relaxation explains why $P_e$ does not saturate to exactly 1 when 3 or 4 levels are populated after the end of the Rabi $\pi$ pulse. We also remark that, though levels $\left| 2 \right>$ and $\left| 3 \right>$ undergo fast relaxation during the flux ramping, their detection probability is not so different from the pure tunnelling prediction given in the first part of the article. In fact, to give a crude picture, the populations relax towards $\left| 1 \right>$ where they accumulate and can be detected.

The discrepancy which appears for small $a_{\mu w}$ is due to decoherence: low amplitude $\mu$w implies a slow Rabi pulse and bigger errors due to incoherent processes.  In a two level system, decoherence during a $\pi$ pulse leads to a population reduction $\tfrac{1}{2} \big[ 1-\exp(-t_{\mu w}/\tilde{T}_2) \big]$ for the first excited state. We use this result to roughly estimate the complex effect of decoherence in the multilevel regime. The previous formula leads to the error bars on the predicted escape probability, plotted in Fig. \ref{fig7}(a).

\section{Conclusion}

In conclusion, we have demonstrated a very fast method for detecting the state of a current biased dc SQUID. This technique exploits an adiabatic but ultrafast reduction of the barrier height seen by the trapped levels, which allows a selective escape of the excited states. Practically, the measurement is performed with a roughly trapezoidal flux pulse whose plateau duration is a few nanoseconds. The detection probability of the ground state is well understood using the standard macroscopic quantum tunnelling formula. We report a slight broadening of the escape curve due to
current noise generated by the close electrical environment of the SQUID. This broadening is by far insufficient to account for the observed experimental contrast of detection in the two level limit. A detailed study of depolarization processes reveals that spurious relaxation during the flux ramping dominates depolarization. The simple model that we have developed explains the observed contrast of multilevel Rabi oscillations, when the measuring pulse amplitude is
set to maximise the detection contrast between $\left| 0 \right>$ and $\left| 1 \right>$. Measurement then gives the population of excited states with a good fidelity. Detection contrast between $\left| n \right>$ and $\left| n +1 \right>$ with $n \geq 1$, however, is strongly lowered by relaxation. Depolarization during the flux ramping thus appears as a clear limitation to the experimental determination of the levels occupancies in the multilevel regime. In phase Josephson circuits, this relaxation is attributed to spurious two level fluctuators, localized inside the dielectric of the Josephson junctions. The remarkable results obtained in Ref. \cite{Steffen_PRL06} allows for optimism. A new design of the dc SQUID, combined with a new dielectric for the junction should lead to a significant improvement of the detection contrast.

We thank W. Guichard, F. W. J. Hekking, L. P. L\'evy, and A. Zazunov for fruitfull discussions. This work was supported by two ACI programs, by the EuroSQUP project and by the Institut de Physique de la Mati\`ere Condens\'ee.

\section*{Appendix 1: from 2D to 1D potential}

This analytical calculation is strongly inspired by Ref. \cite{Lefevre_PRB92}, in the general case where the inductance asymmetry coefficient $\eta$ is finite. Let us consider the two dimensional potential:
\begin{equation}
U(x,y)=2 E_J [-sx-\cos x \cos y - s \eta y + b(y-y_B)^2],
\end{equation}
where $s=I_b/2I_0$ and $y_B=\pi(\Phi_b/\Phi_0)$ are the reduced bias parameters. For the bias points used in the article, the potential surface displays equivalent local minima separated by saddle points. The critical current of the SQUID is geometrically defined as the bias current $I_c$ where the minimum vanishes and merges with adjacent saddle point at the coordinate ($x_c,y_c$). Experiments are performed close to the critical current, therefore a Taylor expansion of the potential in the vicinity of the critical point $(s_c,x_c,y_c)$ is justified. Following \cite{Lefevre_PRB92}, we introduce $\Delta x=x-x_c$ et $\Delta y=y-y_c$. Limiting calculation to the third order one obtains:
\begin{equation}
\begin{split}
U/(2 E_J) &= (s_c-s)(\Delta x + \eta \Delta y)\\
&+ \tfrac{1}{2}\cos x_c \cos y_c \left( \Delta x^2 + \Delta y^2 \right)\\
&- \sin x_c \sin y_c \Delta x \Delta y + b\Delta y^2 \\
&- \tfrac{1}{6} \sin x_c \cos y_c \left( \Delta x^2+3\Delta y^2 \right) \Delta x\\
&- \tfrac{1}{6} \sin y_c \cos x_c \left( \Delta y^2+3\Delta x^2 \right) \Delta y.
\end{split}
\end{equation}
We are interested in the dynamics along the escape direction which corresponds to the direction of minimum curvature at the critical point. It makes an angle $\theta$ with the $x$ axis:   
\begin{equation}
\tan \theta = - \frac{\partial_x^2 U(x_c,y_c)} { \partial_x \partial_y U(x_c,y_c)} = \frac{\cos x_c \cos y_c}{\sin x_c \sin y_c}.
\end{equation}
Let $\phi$ be the phase along the escape direction. Looking for extrema of the potential along this direction, one finds a minimum located at $\phi=0$ and a saddle point. Along the escape direction, the anharmonic potential reads:
\begin{equation}
U(\phi)=m\omega_p^2 \Bigg[ \frac{\phi^2}{2} - \sqrt{\frac{m \omega_p^2}{6 \Delta U}} \frac{\phi^3}{3} \Bigg].
\end{equation}
It is fully determined by the plasma frequency $\nu_p=\omega_p/(2\pi)$ and the barrier height $\Delta U$:
\begin{align}
\Delta U &= \Big( \frac{s_c}{u_3} \Big)^{1/2}\left( \cos \theta + \eta \sin \theta \right)^{3/2} \Delta U^{JJ}(I_b,I_c) \\
\nu_p &= \Big( \frac{u_3}{s_c} \Big)^{1/4}\left( \cos \theta + \eta \sin \theta \right)^{1/4} \nu_p^{JJ}(I_b,I_c,2C_0).
\end{align}
The prefactor $u_3=\sin x_c \cos y_c \cos \theta ( 1+2\sin^2 \theta)+ \sin y_c \cos x_c \sin \theta ( 1+2\cos^2 \theta)$ is the third derivative of the potential along the escape direction. $\Delta U^{JJ}$ and $\nu_p^{JJ}$ are the well known barrier height and plasma frequency for a single Josephson junction with critical current $I_c$, capacitance $2C_0$, biased with a current $I_b$:
\begin{align}
\Delta U^{JJ}(I_b,I_c) &= \Big( \frac{4\sqrt{2}}{3}\frac{\Phi_0}{2\pi}I_c \Big) \big( 1-I_b/I_c \big)^{3/2} \\
\nu_p^{JJ}(I_b,I_c,2C_0) &= \Big( \frac{\sqrt{2}}{2\pi} \frac{1}{\Phi_0}\frac{I_c}{2C_0} \Big)^{1/2} \big( 1-I_b/I_c \big)^{1/4}.
\end{align}

\section*{Appendix 2: coupling to a flux signal}

In the general case, a small flux perturbation $\delta \Phi (t)$ added to the bias flux introduces a perturbation term:
\begin{equation}
\hat{W}(t)=- \frac{\Phi_0}{\pi} \frac{1}{L_s} \delta \Phi (t) \times \hat{y}.
\end{equation}
The component of this operator on the escape direction reads:
\begin{equation}
\hat{W}_{\varphi}(t) = - \frac{\sin \theta}{2\pi} \frac{1}{L_s} \Big( \frac{2 h}{C_0 \nu_p} \Big)^{1/2} \delta \Phi(t) \times \hat{\varphi}.
\end{equation}
Assuming a monochromatic perturbation $\delta \Phi (t)=\Phi_{\mu w} \cos(2\pi \nu t)$, one obtains the perturbation term:
\begin{equation}
\hat{W}_{\varphi}(t) = -  h a_{\mu w} \sqrt{2} \cos(2\pi \nu t) \hat{\varphi},
\end{equation}
where $a_{\mu w}$ is the 2-level Rabi frequency for a tuned excitation ($\nu=\nu_{01}$):
\begin{equation}
a_{\mu w} = \frac{\sin \theta}{2 \pi} \frac{1}{L_s} \frac{1}{\sqrt{C_0 h \nu_p}} \Phi_{\mu w}.
\end{equation}


\begin{thebibliography}{}
\bibitem{RQ_minima} There simutaneously exists two distinct flux states near $\Phi_b/\Phi_0 = 0.5 + n \pi$ (with n a relative integer), with associated non-equivalent minima and critical current. These particular bias points are not studied in this article.
\end{thebibliography}
\end{document}